\documentclass[prb,aps,twocolumn,groupedaddress,floats,showpacs,final,superscriptaddress,10pt]{revtex4-2}
\usepackage[utf8]{inputenc}
\usepackage[T1]{fontenc}
\usepackage{lmodern}
\usepackage[makeroom]{cancel}
\usepackage{graphicx}
\usepackage{amsmath}
\usepackage{amsfonts}
\usepackage{amssymb}
\usepackage{chngcntr}
\usepackage{color}
\usepackage{cancel}
\usepackage{mathtools}
\usepackage[left=2cm,right=2cm,top=2cm,bottom=2cm]{geometry}

\usepackage{graphicx}
\usepackage{dcolumn}
\usepackage{bm}
\usepackage{simplewick}
\usepackage{array}
\usepackage{appendix}

\RequirePackage[
   hyperindex,colorlinks,bookmarksnumbered,
   plainpages=true,pdfstartview=FitH]{hyperref}
\hypersetup{linkcolor=blue,urlcolor=blue,citecolor=blue}

\definecolor{purple}{rgb}{0.5,0,0.6}

\usepackage{ulem}

\begin{document}

\title{
Conductivity of charge-neutral multicomponent 2D electron-hole system}

\author{Yuping Huang}
\affiliation{Department of Materials Science and Engineering, Southern University of Science and Technology, 1088 Xueyuan Blvd, Shenzhen, 518055, China}
\affiliation{Guangdong Technion -- Israel Institute of Technology, 241 Daxue Road, Shantou, Guangdong, China, 515063}


\author{O.~V.~Kibis}
\affiliation{Novosibirsk State Technical University, Novosibirsk 630073, Russia}

\author{V.~M.~Kovalev}
\affiliation{Novosibirsk State Technical University, Novosibirsk 630073, Russia}


\author{I.~G.~Savenko}
\email{ivan.g.savenko@gmail.com}
\affiliation{Guangdong Technion -- Israel Institute of Technology, 241 Daxue Road, Shantou, Guangdong, China, 515063}
\affiliation{Guangdong Provincial Key Laboratory of Materials and Technologies for Energy Conversion, Guangdong Technion--Israel Institute of Technology, Guangdong 515063, China, 515063}

\date{\today}

\begin{abstract}
The interplay between distinct carrier species in systems with broken Galilean invariance can give rise to a rich landscape of interaction-driven transport phenomena. 
Here, we develop a comprehensive theory of the electrical conductivity of a two-dimensional mixture of massless Dirac and massive fermions, a system realized in HgTe quantum wells tuned to the charge-neutrality point. 
In this regime, all carriers are thermally activated, enabling a self-consistent, temperature-dependent interplay between the two species. Crucially, the charge-neutrality condition ensures that the chemical potential is not externally pinned but is determined self-consistently, thereby making the system's transport response an intrinsic probe of inter-species quantum friction. 
\textcolor{black}{We demonstrate that the conductivity changes with temperature :} at low temperatures, transport is dominated by massless Dirac carriers, yielding a nearly temperature-independent conductivity reminiscent of the charge-neutrality point in pristine graphene. 
As the temperature rises, massive holes become thermally excited, and their mutual scattering with Dirac carriers induces a specific nonmonotonic temperature dependence of the conductivity in both clean and disordered structures. 
In particular, in nearly clean structures with strong screening modeled by a short-range interparticle interaction potential, the system's conductivity can exhibit an inverse-quadratic temperature dependence. 
Conversely, in disordered structures with a long-range interparticle interaction, it varies quadratically with temperature. 
Our findings establish HgTe quantum wells at charge neutrality as a clean, highly tunable platform for isolating and quantitatively studying interaction-driven transport in the absence of Galilean invariance, offering a direct pathway to explore regimes where interparticle collisions dominate over disorder.
\end{abstract}

\maketitle

\section{Introduction}
The temperature dependence of the conductivity of non-Galilean-invariant two-dimensional (2D) systems exhibits unique features~\cite{Novoselov2004, Skakalova2009}. 
Unlike conventional semiconductors with parabolic carrier dispersion, where Galilean invariance ensures momentum conservation and thus prevents interparticle interactions from contributing directly to DC resistivity, this constraint is lifted in non-Galilean-invariant 2D systems. 
This enables electron-electron collisions to significantly alter or even dominate charge transport at moderate temperatures~\cite{Pal2012Resistivity}.
In particular, for pristine (and gapless) graphene at the charge neutrality point (CNP), the dominant mechanism is electron-hole (e-h) scattering, leading to temperature-independent conductivity (without direct accounting for phonon scattering)~\cite{VyurkovRyzhii, PhysRevB.78.085415, PhysRevB.78.085416, Levin2026}.
In the presence of a band gap, the temperature dependence of the gapped graphene conductivity exhibits exponential behavior that obeys the Arrhenius law, arising from the combined effects of the gap opening and impurity scattering. 
Conversely, pristine gapped graphene exhibits a linear temperature dependence of the conductivity~\cite{PhysRevB.109.085424}.


The development of growth techniques for producing clean 2D semiconducting materials, including transition-metal dichalcogenides and HgTe-based quantum wells (QWs), opens new perspectives on controlled particle transport in 2D. 
In particular, at low temperatures, 2D systems might exhibit a regime where interparticle collisions dominate over impurity- or phonon-mediated scattering. 
For example, thermally excited electron-hole pairs in tunable twisted bilayer graphene semimetal can provide the main contribution to conductivity~\cite{NamNatPhys2017, doi:10.1126/sciadv.abi8481}, yielding $T^2$-scaled resistivity~\cite{PhysRevLett.129.206802}. 
In transition metal dichalcogenides, interparticle interactions may also induce a non-vanishing correction to the longitudinal and transverse conductivities. This contribution originates from intervalley scattering, as was theoretically demonstrated~\cite{PhysRevB.109.245414, PhysRevB.110.L041301, PhysRevB.106.235305}. 

These examples of unique transport responses have spurred interest in identifying other 2D platforms with broken Galilean invariance, in which diverse carrier dispersions (linear, parabolic, or mixed) may induce novel interaction-driven phenomena.
In particular, semimetals like HgTe-based structures can serve as an example of such a system exhibiting fascinating transport properties:
They support two distinct charge carriers, electrons and holes, with different energy dispersions, which enables the breaking of Galilean invariance when both particle gases are sufficiently dense (i.e., in the degenerate electron and hole gas regime)~\cite{PhysRevB.67.115316, PhysRevB.102.155411, PhysRevB.106.085411}. 
Consequently, electron-hole friction arises, contributing measurably to resistivity~\cite{PhysRevLett.95.146802}.

\begin{figure}[t!]
\includegraphics[width=0.8\columnwidth]{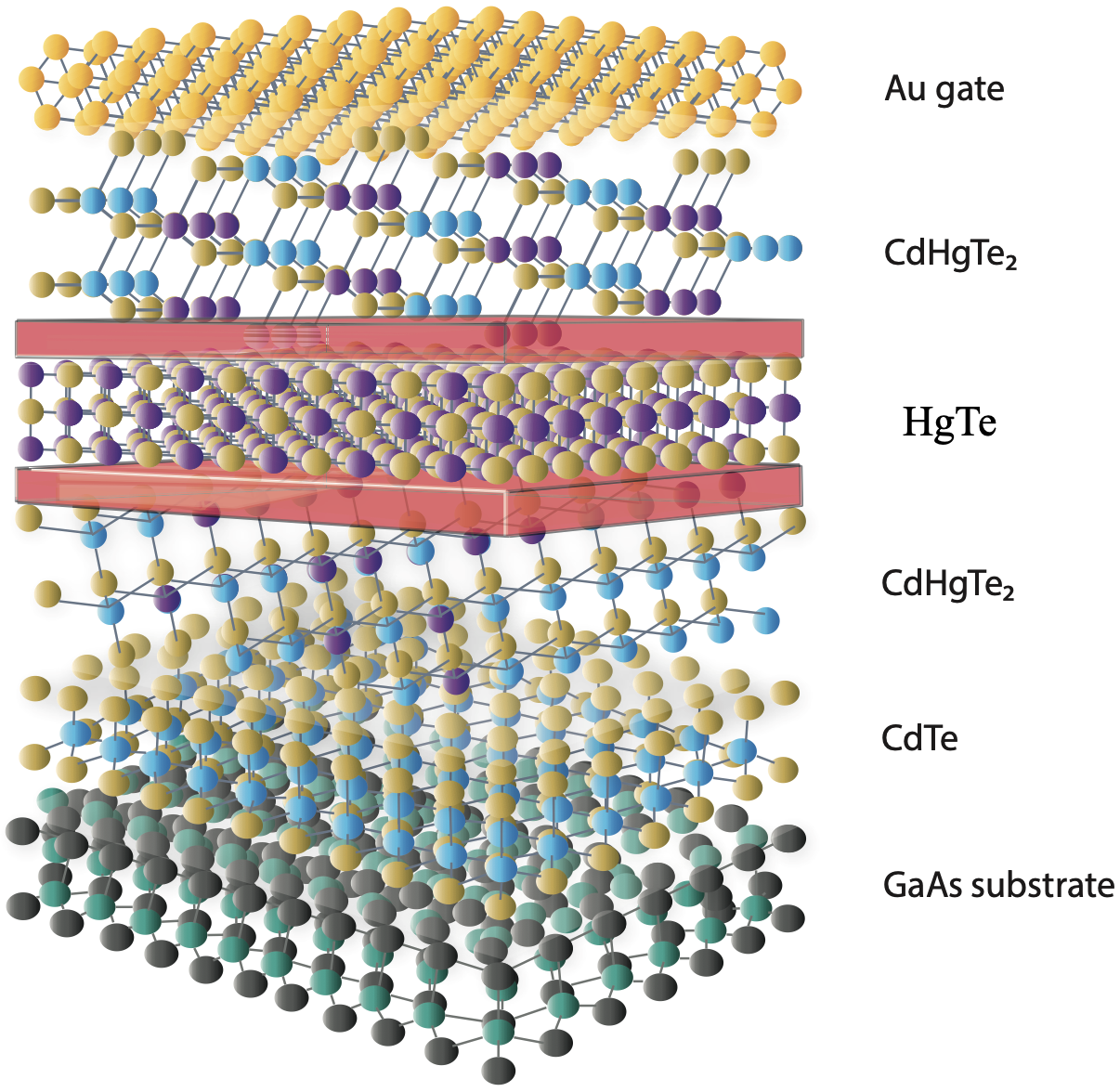} 
\caption{A typical experimental heterostructure: a 2D electron and hole gases of Dirac particles and massive particles in a HgTe QW sandwiched between other layers.}
\label{Fig1}
\end{figure}

Generally, HgTe-based QWs sandwiched between other layers (Fig.~\ref{Fig1}) constitute a uniquely versatile platform, whose properties are defined by two critical parameters. 
The first of them is fixed during the growth process, the QW width. 
It dictates the energy spectrum~\cite{PhysRevLett.95.226801}, and near a critical value ($\sim$6 nm), the bandgap vanishes, enabling the transition to 2D topological insulators~\cite{doi:10.1126/science.1148047, GUSEV2019113701} or semimetals~\cite{EntinJETP2013, ButtnerNatPhys2011}, which are the states characterized by a single Dirac cone valley and distinctive transport behavior such as magnetotransport and the quantum Hall effect~\cite{KozlovHETPL2013, KvonJETPL2008, PhysRevB.96.045304, PhysRevB.99.121405, nano12142492}. 
The second parameter, the gate voltage, is tunable in situ: it controls the carrier density and degeneracy, allowing switching between regimes (e.g., from a topological insulator to a semimetal).
Magnetic fields further expand the potential tunability and phenomenology of these systems: they break symmetries to induce unconventional transport phenomena (e.g., asymmetric electron-phonon interactions in quasi-2D structures~\cite{Lawton-2002, Kibis_1999, Kibis_2001, Kibis_2002}) while acting as a tool for manipulating material states.

In our recent work~\cite{Huang2026}, we studied the effect of carrier--carrier interactions on magnetotransport in p-doped HgTe QWs in the degenerate regime~\cite{Huang2026}. 
There, the system hosts both massless Dirac holes and massive holes, with the chemical potential lying above the energy gap, $\mu > \Delta$, where $\Delta$ is the energy gap of the massive-hole subsystem, measured from the Dirac point. 
The analysis showed that when both the massive and Dirac holes coexist, finite interaction-mediated corrections appear. 
These scale as $T^2$ for short-range interactions and $T^2\ln(1/T)$ for unscreened Coulomb interactions; both are suppressed in the case of a strong magnetic field.

That study was restricted to the degenerate electron and hole gas regime. 
A complementary non-degenerate regime, in which some carrier densities are low enough for Boltzmann statistics to apply, has not been addressed within the same framework. This regime is relevant at higher temperatures, 
with the chemical potential set self-consistently by charge neutrality. \textcolor{black}{Moreover, exactly this regime has been recently realized experemintally~\cite{Levin2026}.}

\begin{figure}[b!]
\includegraphics[width=0.99\columnwidth]{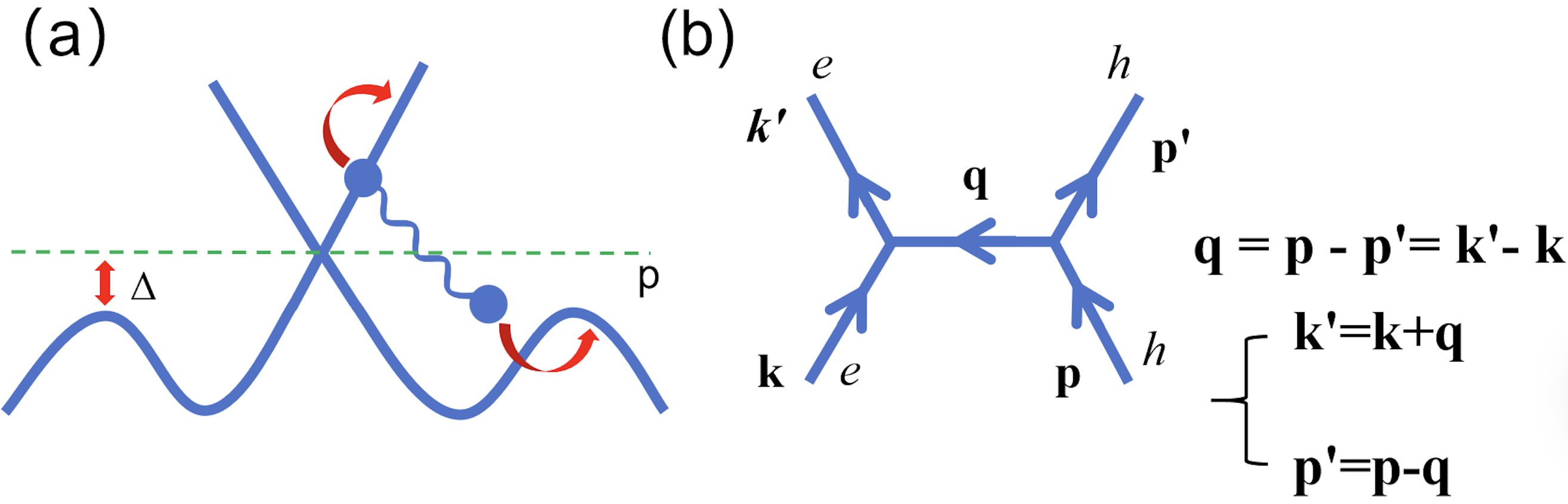} 
\caption{Schematic of the most important scattering processes in the system: the scattering of Dirac electrons on massive holes and vice versa, as visualized on the dispersion spectrum in (a) and the respective Feynman diagram in (b).
The electrons are spread over a linear Dirac-like spectrum, and the holes spectrum consists of a linear-in-momentum domain at low momenta $p$, and a quadratic-in-momentum part in the vicinity of $p\neq 0$.}
\label{Fig2}
\end{figure}

In this manuscript, we consider the case of a 2D system acquiring charge carriers with the dispersion presented in Fig.~\ref{Fig2}(a).
Thus, we consider a similar HgTe-based QW system, but now operating in the charge-neutrality regime, where the total electron density precisely balances the sum of the densities of the two distinct hole species. 
In this regime, the massive holes are thermally activated across the energy gap $\Delta$, gradually replacing the contribution from massless holes as temperature increases.
Crucially, the chemical potential is not pinned by external doping; instead, it is determined self-consistently from the charge neutrality condition as a function of temperature.

Applying linear response theory to a weak electric field, we employ the linearized Boltzmann transport equation to calculate the interparticle-collision corrections to the electrical conductivity. The collision integral in this formulation explicitly accounts for the Coulomb scattering processes, both within each carrier species (intra-species) and between carriers from different species (inter-species). 
This approach fully captures the interaction-driven dynamics inherent in this hybrid-carrier system.
\begin{figure*}[t!]
\includegraphics[width=1.9\columnwidth]{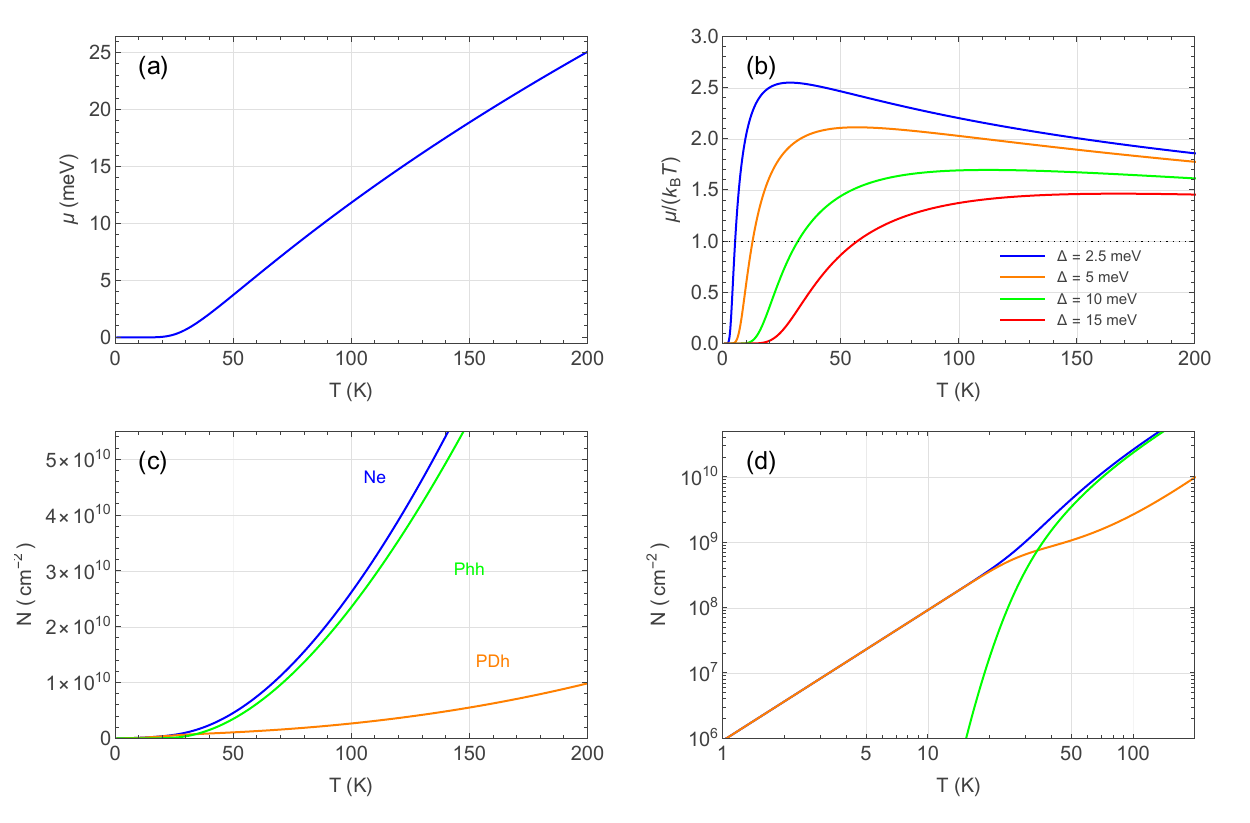} 
\caption{Temperature dependence of (a) the chemical potential, (b) the degeneracy factor $\mu/T$ \textcolor{black}{for different values of $\Delta$}, and (c, d) the concentrations of Dirac electrons (e), Dirac holes (Dh), and massive holes (hh) plotted on (c) linear and (d) double-logarithmic scales, as found in the numerical solution of the neutrality equation. 
In calculations, we used the following parameters: $v = 7.0 \times 10^5~\mathrm{m/s}$, 
$m = 0.15\,m_0$, and $\Delta=15$~meV for panels (a), (c) and (d). 
}
\label{Fig3}
\end{figure*}

\textcolor{black}{
We show that up to 50~K (for realistic parameters of HgTe QWs), the conductivity behaves like that of graphene.
It represents a symmetric two-component plasma of massless Dirac electrons and holes.
Then, massive holes with a gap get activated, and the system enters a complex, moderate-temperature crossover regime.
Here, the population of thermally activated massive holes grows, shifting the chemical potential. 
At higher temperatures ($T>50$~K), the system transitions entirely into a hybrid plasma of moderately degenerate Dirac electrons and non-degenerate massive Boltzmann holes. 
The conductivity demonstrates a power-law behavior as a function of temperature.
Here, interparticle collisions and scattering on disorder coexist and compete with each other. 
At even higher temperatures, phonon scattering dominates. 
However, this regime is beyond the scope of our consideration (as it has been addressed in previous works on classical semiconductors).
}



\section{Equilibrium properties}
The equilibrium distribution functions of the particles read as
\begin{eqnarray}
n_{e(Dh)}({\bf k})=\frac{1}{e^{(\varepsilon_{\bf k}\mp\mu)/T}+1},
\\
n_{hh}({\bf p})=\frac{1}{e^{(\varepsilon_{\bf p}+\Delta+\mu)/T}+1},
\end{eqnarray}
where $\varepsilon_{\bf k}=vk$ and $\varepsilon_{\bf p}={\bf p}^2/2m$ are the kinetic energies of Dirac carriers [electrons ($e$) and holes ($Dh$)] and massive holes ($hh$) counted from  $\Delta$, respectively. 
Here and below, the momentum $\bf k$ denotes the dispersion of Dirac particles, while $\bf p$ denotes the dispersion of the massive hole sector.
Then, the equilibrium particle densities read as
\begin{gather}
\label{EqDiffDens01}
N_e=\sum_{\bf k}n_e({\bf k})=-\frac{T^2}{2\pi v_F^2} \text{Li}_2\left(-e^{\mu/T}\right)
,\\
\nonumber
P_{Dh}=\sum_{\bf k}n_{hD}({\bf k})=-\frac{T^2}{2\pi v_F^2} \text{Li}_2\left(-e^{-\mu/T}\right)
,\\
\nonumber
P_{hh}=\sum_{\bf p}n_{hh}({\bf p}) = \frac{m T}{2\pi} \ln \left( 1 + e^{-\frac{\Delta+\mu}{T}} \right),
\end{gather}
where
\begin{equation}
\text{Li}_2\left(-e^{y}\right) = -\int\limits_{0}^{\infty} \frac{x}{e^{x-y} + 1} \, dx
\end{equation}
is the polylogarithm function of the second order. 

The electroneutrality equation, which is fundamental for any semiconductor at 
charge neutrality, is given by $N_e = P_{\text{Dh}} + P_{\text{hh}}$. 
This relation 
determines the temperature dependence of both the degeneracy factor, $\mu(T)/T$, and the respective carrier densities.
An exact analytical treatment of this equation is not possible, so numerical analysis is required to gain further insight.

Figure~\ref{Fig3} displays the calculated temperature dependences of the chemical 
potential, the degeneracy factor, and the particle densities. 
As the temperature increases, the system undergoes significant shifts in the carrier populations. 
At sufficiently low temperatures, the massive-hole density remains negligible, and the system behaves as a compensated Dirac semimetal with equal electron and hole concentrations. 
This regime is closely reminiscent of gapless graphene, the conductivity of which 
has been intensively studied. 
A further thermal excitation activates the massive holes; their density 
eventually surpasses that of the Dirac holes, while the Dirac hole concentration subsequently adjusts to satisfy the charge neutrality condition.


\section{Analytical model of equilibrium properties}
\textcolor{black}{Here, we focus exclusively on the two-component plasma regime within the moderate-temperature range, where the massive-hole density dominates over the Dirac-hole density. 
According to the numerical calculations presented in the previous section, this corresponds to the temperature range (approximately) \(T = 50\div150\)~K for realistic HgTe structures. 
The theoretical interest in this regime is strongly motivated by recent experimental data~\cite{Levin2026}, which were obtained precisely when the massive holes play a dominant role.
Furthermore, the two-component description enables the construction of an analytical framework that yields simple theoretical expressions describing the system's behavior in this moderate-temperature regime.
}

Notably, in the moderate-temperature range, where the Dirac hole density becomes negligible, the chemical potential is positive and increases with temperature. 
This causes the degeneracy factor to reach the `degeneracy plateau' with $\mu/T > 1$, indicating the moderate degeneracy of the Dirac electron component. 
Concurrently, the massive-hole component obeys deep Boltzmann statistics due to the large energy factor $\Delta + \mu > T$. 
Then, the chemical potential shifts upward deep into the conduction band, enforcing inequality $\mu/T > 1$. 

\textcolor{black}{Furthermore, an analytical theory can be built assuming $\mu/T\gg1$. Thus, } we can establish the following asymptotic simplifications:

i) For the Dirac electrons, we use the low-temperature expansion:
\begin{equation}\label{DiracElectronDens}
N_e \approx \frac{\mu^2}{4\pi v^2} + \frac{\pi T^2}{12 v^2}. 
\end{equation}
%
We employ Eq.~\eqref{DiracElectronDens} to further develop a transparent analytical treatment of the problem, \textcolor{black}{keeping in mind that this approximation is as accurate as the inequality $\mu>T$ is stronger.}

ii) For the density of intrinsic Dirac holes, they freeze out exponentially due to the high Fermi barrier, rendering their contribution to the net charge balance negligible:
\begin{equation}
P_{Dh} \approx \frac{T^2}{2\pi v^2} e^{-\mu/T} \to 0.
\end{equation}

iii) Due to the high combined activation barrier $\Delta + \mu > T$, the massive holes enter the non-degenerate regime and perfectly obey classical Maxwell-Boltzmann statistics:
\begin{equation}
P_{hh} \approx \frac{m T}{2\pi} \exp\left(-\frac{\Delta + \mu}{T}\right).
\end{equation}
Thanks to dropping out the frozen-out Dirac holes, we come up with an actual hybrid charge neutrality equation $N_e\approx P_{hh}$, yielding:
\begin{equation}
\label{eq:hybrid}
\frac{\mu^2}{4\pi v^2} + \frac{\pi T^2}{12 v^2} \approx \frac{m T}{2\pi} \exp\left(-\frac{\Delta + \mu}{T}\right).
\end{equation}

It would be useful to have an analytical solution to Eq.~\eqref{eq:hybrid} in the moderate-temperature regime. 
To solve~(\ref{eq:hybrid}) in the leading temperature order, we omit the second-order correction $\sim T^2$. 
Then, transforming the equation into the canonical form $we^w = z$, where $w=\mu/2T$ and $z=\exp(-\Delta/2T)\sqrt{mv^2/2T}$, enables us to express the temperature trajectory of the chemical potential through the principal branch of the Lambert function $W_0(z)$ as $w=W_0(z)$, yielding:
\begin{equation}
\label{eq:mu_lambert}
\mu(T)=2T W_0\left(\sqrt{\frac{m v^2}{2T}} e^{-\Delta/(2T)}\right).
\end{equation}
This expression qualitatively reproduces the numerically exact behavior of the degeneracy factor shown in Fig. \ref{Fig3}(b), albeit yielding slightly higher values.

Furthermore, using the relation $e^{-W_0(z)} = W_0(z)/z$, and substituting the explicit chemical potential~(\ref{eq:mu_lambert}) into the Boltzmann distribution, the massive hole density yields:
\begin{eqnarray}
\label{HHdensity}
P_{hh} &\approx& \frac{m T}{2\pi} e^{-\Delta/T} e^{-2W_0(z)} \\
\nonumber
&&~~~~~=\frac{m T}{2\pi} e^{-\Delta/T} \left( \frac{W_0(z)}{z} \right)^2=\frac{T^2}{\pi v^2} W_0^2(z).
\end{eqnarray}
Concurrently, for the Dirac electrons in the leading order of degeneracy, we find:
\begin{equation}
N_e \approx \frac{\mu^2}{4\pi v^2} = \frac{(2T W_0(z))^2}{4\pi v^2} = \frac{T^2}{\pi v^2} W_0^2(z).
\end{equation}
%
%
These formulas explicitly demonstrate that within the degeneracy factor plateau domain, corresponding to the moderate temperature regime, the densities of both the participating subsystems $N_e\approx P_{hh}$ scale according to a quadratic law in $T$: $N_e\approx P_{hh}\sim T^2 / (\pi v^2)$, modulated by the dimensionless factor $W_0^2(z)$. 


\section{Kinetics of interacting Dirac electrons -- massive holes system}
As was stated above, we focus on
the moderate-temperature regime, where Dirac electrons and massive holes dominate.
Furthermore, we distinguish between two main interparticle collision mechanisms that contribute to the electric current density: the scattering of Dirac electrons off massive holes and the scattering of massive holes off Dirac electrons, while disregarding other possible scattering processes, which are usually weaker.

The Boltzmann transport equations describing the scattering of massless electrons on impurities and massive holes, and, accordingly, the scattering of massive holes on electrons and impurities, read as~\cite{PhysRevB.109.245414} (here and below, we use $c=\hbar=1$ units in the derivations and later restore the correct dimensionality)
\begin{eqnarray}
\label{EqElectrons00}
-({\bf F}
\cdot\nabla_{\bf k})f_e({\bf k})+\frac{f_e({\bf k})-n_e({\bf k})}{\tau(\varepsilon_{\bf k})}=
Q_{eh}+Q_{ee},~~~\\
\nonumber
({\bf F}
\cdot\nabla_{\bf p})f_{hh}({\bf p})+\frac{f_{hh}({\bf p})-n_{hh}({\bf p})}{\tau(\varepsilon_{\bf p})}=Q_{he}+Q_{hh},
\end{eqnarray}
where $e>0$ is the elementary charge, ${\bf F}=e{\bf E}$, $\mathbf{E}$ is the electric field,  
$f$ and $n$ are the nonequilibrium and equilibrium distribution functions, respectively, and $Q_{eh(he)}$ and $Q_{ee(hh)}$ are the electron-hole ($eh$), electron-electron ($ee$) and hole-hole ($hh$) collision integrals. The particle scattering off impurities is accounted for via the relaxation-time approximation.


\subsection{Impurity scattering}
Due to distinct density of states for different carrier types, the momentum relaxation times differ for short-range and long-range (Coulomb) models of scattering centers~\cite{RevModPhys.83.407}.
The respective momentum relaxation times for Dirac electrons read as
\begin{equation}\label{diractimes} 
\frac{1}{\tau_e(\varepsilon_{\bf k})}= 
\begin{dcases} 
\frac{N_i|u_0|^2\varepsilon_{\bf k}}{4\hbar^3v^2}=\dfrac{1}{4\tau_i}\dfrac{\varepsilon_{\bf k}}{mv^2}, & \text{short-range}, \\ 
\frac{(2\pi e^2)^2N_i}{4\epsilon^2\hbar\varepsilon_{\bf k}}=
\dfrac{1}{\tau_\textrm{C}}\dfrac{mv^2}{\varepsilon_{\bf k}}, & \text{long-range}, 
\end{dcases} 
\end{equation}
whereas for massive holes
\begin{equation}\label{heavytimes}
\frac{1}{\tau_h(\varepsilon_{\bf p})}= 
\begin{dcases}
\dfrac{1}{\tau_i},   & \text{short-range}, \\
\frac{(2\pi e^2)^2N_i}{4\epsilon^2\hbar\varepsilon_{\bf p}}=
\dfrac{1}{\tau_\textrm{C}}\dfrac{mv^2}{\varepsilon_{\bf p}},  & \text{long-range}.
\end{dcases}
\end{equation}
Here $1/\tau_i=mN_i|u_0|^2/\hbar^3$ and $1/\tau_\textrm{C}=(2\pi e^2/\epsilon\hbar v)^2N_i\hbar/4m$ are parameters describing the strength of the impurity centers' interaction with mobile carriers, and $N_i$ is the density of impurity centers. 
Below, the relaxation times~\eqref{diractimes} and~\eqref{heavytimes} are used to investigate the temperature-dependent behavior of the system's conductivity.


\subsection{Electron-hole scattering in the moderate temperature range}
Owing to the two-component nature of the system, interparticle scattering is governed by two distinct mechanisms: intra-subsystem scattering (within individual Dirac electron and massive hole subsystems) and inter-subsystem electron-hole scattering. 
Notably, hole-hole scattering does not contribute to transport relaxation. Indeed, due to the Galilean invariance of the hole subsystem, which features a parabolic dispersion, hole-hole collisions conserve both total momentum and velocity, thereby yielding no net electrical resistance. 

Conversely, both Dirac electron-electron collisions and electron-hole scattering explicitly contribute to the resistivity. At the charge-neutrality point of graphene, intra-band electron-electron scattering processes are weak in electrical conductivity due to severe kinematic restrictions imposed by the linear Dirac spectrum. Consequently, the finite electrical resistivity of graphene is heavily dominated by electron-hole friction. Following these established models~\cite{PhysRevB.78.085415, PhysRevB.78.085416}, we likewise restrict our consideration exclusively to the effects of Dirac electrons-massive holes scattering on the transport properties of the system.

Let us consider the electron-hole collision integrals, $Q_{eh(he)}$, in the kinetic equations for Dirac electrons and massive holes. 
Expressing the nonequilibrium corrections to the electron and hole distribution functions in the standard form (using the standard ansatz~\cite{Zaitsev:2014aa}), we find
\begin{eqnarray}
\label{nregime4}
\delta f_{\bf k}&=&f_e({\bf k})-n_e({\bf k})
\equiv\chi_e({\bf k})n_e'({\bf k}),
\\
\delta f_{\bf p}&=&f_{hh}({\bf p})-n_{hh}({\bf p})
\equiv\chi_h({\bf p})n_{hh}'({\bf p}),
\end{eqnarray}
where the prime sign means the derivative with respect to energy. 
Then, the electron-hole collision integral can be linearized, yielding:
\begin{gather}\nonumber
Q_{eh(he)}=2\pi\sum_{\mathbf{l},\mathbf{k}',\mathbf{p}',\mathbf{q}}
|U_{\mathbf{k}'-\mathbf{k}}|^2
{\cal F}_{\bf kk'}\\\nonumber\times
[\chi_h(\mathbf{p})-\chi_h({\mathbf{p}'})+\chi_e(\mathbf{k})-\chi_e({\mathbf{k}'})]
\\\nonumber
\times[n_{hh}(\mathbf{p})-n_{hh}({\mathbf{p}'})][n_e(\mathbf{k})-n_e({\mathbf{k}'})]
\delta_{\mathbf{k}',\mathbf{k}+\mathbf{q}}
\delta_{\mathbf{p}',\mathbf{p}-\mathbf{q}}\\
\label{EqCollisionIntegral00}
\times
\int \frac{d\omega \delta(\varepsilon_{{\bf k}'}-\varepsilon_{\bf k}-\omega)
\delta(\varepsilon_{{\bf p}'}-\varepsilon_{\bf p}+\omega)}{(-4T)\sinh^2(\frac{\omega}{2T})},
\end{gather}
where ${\bf q}$ and $\omega$ are the momentum and energy, which are transferred between the particles in the collision event, and the summation over ${\bf l}$ means ${\bf l}={\bf k}$ for $Q_{he}$ and ${\bf l}={\bf p}$ for $Q_{eh}$; ${\cal F}_{\bf kk'}=(1+\cos\theta_{\bf kk'})/2$.

A principal simplification, which can be applied to this expression, is the assumption that the electron-hole scattering processes are nearly elastic. 
Indeed, in an individual scattering event, an electron transfers energy $\omega$ and momentum $\mathbf{q}$ to the hole subsystem. 
The exact energy conservation law for electrons stemming from Eq.~\eqref{EqCollisionIntegral00} reads as $\omega = \sqrt{\mu^2 + (q v)^2 + 2 \mu q v \cos\phi} - \mu$, where $\phi$ is the angle between the initial electron momentum and the transferred momentum $\mathbf{q}$.
The thermal factors in the collision integral strictly constrain the magnitude of the actually exchanged energy to the scale of the system temperature ($\omega \le T$). 
Squaring the energy conservation law yields a rigorous quadratic equation for the transferred momentum: $(qv)^2 + 2\mu \cos\phi \cdot (qv) - \omega (2\mu + \omega) = 0$. 
The only physically meaningful positive root of this equation is $q(\phi, \omega) = (-\mu \cos\phi + \sqrt{\mu^2 \cos^2\phi + 2\mu \omega + \omega^2})/v$. 
To determine the kinematic boundaries of the transferred momentum, we fix the maximum transferred energy at the level of $\omega = T$ and examine the entire range of scattering angles $\phi \in [0, \pi]$. 

The minimum momentum transfer is achieved for forward scattering ($\phi = 0$), and amounts to $q_{\min} = (-\mu + \sqrt{(\mu + T)^2})/v = T/v$, whereas the maximum momentum transfer occurs for nearly backscattering ($\phi = \pi$), and is strictly bounded by $q_{\max} = \left(\mu+\sqrt{(\mu  + T)^2}\right)/v=(2\mu+T)/v\approx2\mu/v$. 
Thus, without any prior approximations, the absolute value of the transferred momentum is confined within the range $T/v \le q \le 2\mu/v$. 

Let us compare these boundaries with the scale of the intrinsic thermal momentum of the massive holes, which is determined from the thermodynamic equilibrium conditions as $p = \sqrt{2mT}$. 
Dividing the initial inequality by $p$, we find 
$\frac{T}{v\sqrt{2mT}} \le q/p \le \frac{2\mu}{v\sqrt{2mT}}$. 
Bringing the parameters under the radicals and after some algebraic cancellations, we arrive at a rigorous  inequality for the momentum ratio: 
$\sqrt{\frac{T}{2 m v^2}} \le q/p \le \sqrt{\frac{4\mu^2}{2m v^2T}}$. 
\textcolor{black}{Utilizing the commonly introduced auxiliary  `effective mass' of a Dirac electron at the Fermi level~\cite{Levin2026}, $m^* = \mu / v^2$, the upper bound can be rewritten as:} 
\begin{eqnarray}
\sqrt{\frac{T}{2 m v^2}} \le q/p \le \sqrt{2\frac{m^*}{m} \frac{\mu}{T}}.
\end{eqnarray}
In realistic HgTe structures, $m\sim0.15~m_0$ and $ v\sim7\times10^7$~cm/s, yielding $mv^2\sim5\times10^3$~K. 
Thus, for any reasonable temperatures, the left-hand side of this inequality $\sqrt{T/2mv^2}\ll1$. 
Since the massive hole mass is large compared to the effective Dirac electron mass~\cite{Levin2026} ($m \gg m^*, m^*=\mu/v^2\approx10^{-3}m_0,~m=0.15m_0$) and the degeneracy factor can be estimated as $\mu/T\sim2$ according to our numerical analysis, the right-hand side of the inequality is also guaranteed to be small.

\textcolor{black}{This allows us to assume that the momentum $q$ transferred by the electron is small compared to the average thermal momentum of the hole $p$ across the entire range of scattering angles. This justifies the use of the purely elastic scattering approximation for electron scattering off the hole subsystem in the collision integral.
}
Consequently, within this elastic approximation, the impact of electron-hole collisions on the hole subsystem is negligible, whereas their influence on the electron dynamics is of fundamental importance. 
Therefore, {to neglect the influence of electron-hole scattering on hole dynamics, we set} $Q_{he}\approx 0$ (\textcolor{black}{see Supplemental\_Material\cite{SMBG} for the estimation of this contribution and its comparison with the hole-impurity scattering rate}).

Disregarding ${\bf q}$ in comparison with ${\bf p}$, we find a simplified version of the electron-hole scattering integral:
\begin{eqnarray}
Q_{eh}&=&-2\pi n_{e}'(\mathbf{k})\sum_{\mathbf{p},\mathbf{q}}
n_{hh}(\mathbf{p})|U_{\mathbf{q}}|^2
{\cal F}_{{\bf k},\bf {k+q}}\\
\nonumber
&&\times
[\chi_e(\mathbf{k})-\chi_e({\mathbf{k}+{\bf q}})]
\delta(\varepsilon_{{\bf k}+{\bf q}}-\varepsilon_{\bf k}).
\end{eqnarray}
Here, we used the Boltzmann distribution for the massive holes and the degenerate one for the Dirac electron subsystem. 
The quasi-elastic character of the scattering allows us to define the electron-hole relaxation time as $Q_{eh}=-\delta f_{\bf k}/\tau_{eh}(\varepsilon_{\bf k})$, where
\begin{eqnarray}
&&\frac{1}{\tau_{eh}(\varepsilon_{\bf k})}=2\pi P_{hh}\sum_{\mathbf{k'}}
|U_{\mathbf{k}'-\mathbf{k}}|^2
{\cal F}_{\bf kk'}\\
\nonumber
&&~~~~~\times(1-\cos\theta_{\mathbf{k}'\mathbf{k}})\delta(\varepsilon_{{\bf k}'}-\varepsilon_{\bf k}),
\end{eqnarray}
with the hole density $P_{hh}$ depending on temperature according to Eq.~\eqref{HHdensity}. 

The next step is possible if we employ a particular model of interparticle interaction potential $U_{\bf q}$. 
In realistic HgTe structures, the QW has finite width, causing the carriers' transverse wave functions to extend into the surrounding barrier layers. 
Another crucial factor is screening, which significantly modifies the functional forms of both the interparticle and particle-impurity interaction potentials. 
Directly incorporating these effects, however, leads to computationally expensive and cumbersome numerical calculations. 


To circumvent this difficulty and obtain an analytical estimate of the temperature dependence of the system's conductivity, we neglect the penetration of the wave functions into the surrounding layers and employ two simplified models of the interaction potentials that effectively account for screening. 
These models are: (i)~a short-range potential, $U_{\mathbf{q}}=U_0$, which mimics strong screening of the interaction potential and thus describes particle collisions as hard-sphere scattering, and (ii)~a long-range bare Coulomb potential, $U_{\mathbf{q}}=2\pi e^2/\epsilon q$, which accurately accounts for the long-range forces between the colliding particles.
\textcolor{black}{These two limiting cases of the interaction potential are rigorously justified in the Supplemental Material~\cite{SMBG}, where we consider the screening effects induced by the gate and mobile carriers in the system. }

These two cases give
\begin{equation}\label{ehtimes}
\dfrac{1}{\tau_{eh}(\varepsilon_{\bf k})}=
\begin{dcases}
\dfrac{1}{4\tau_0}\dfrac{\varepsilon_{\bf k}}{mv^2}, 
\,\,\,
\dfrac{1}{\tau_0}=P_{hh}\dfrac{m|U_0|^2}{\hbar^3}, \,\,\text{short-range},
\\
\dfrac{1}{\tau_1}\dfrac{mv^2}{\varepsilon_{\bf k}},
\,
\dfrac{1}{\tau_1}=\left(\dfrac{2\pi e^2}{\epsilon\hbar v}\right)^2\dfrac{\hbar P_{hh}}{4m}, \,\,\text{long-range}.
\end{dcases}
\end{equation}
%
%
%
%
The corresponding expression for the Drude conductivity of a nearly degenerate Dirac electron gas is given by:
\begin{gather}
\label{ehcontribution}
\sigma_e = \frac{e^2}{4\pi \hbar} 
\frac{\mu(T)}{\hbar}\left[
\frac{\tau_e(\varepsilon_{\bf k})\tau_{eh}(\varepsilon_{\bf k})}{\tau_e(\varepsilon_{\bf k})+\tau_{eh}(\varepsilon_{\bf k})}
\right]_{\varepsilon_{\bf k}=\mu(T)}.
\end{gather}
This quantity depends on temperature in a non-trivial manner, which arises from both the temperature dependence of the chemical potential and the functional form of the expression within the brackets.

As for the conductivity of the massive hole component, within the elastic approximation, it is primarily determined by hole-impurity scattering processes alone:
\begin{gather}
\label{hhcontribution}
\sigma_{hh}(T)=e^2\frac{P_{hh}(T)}{m}\langle[\varepsilon_{\bf p}\tau_h(\varepsilon_{\bf p})]'\rangle,
\end{gather}
where the prime denotes the derivative with respect to the massive hole energy, and the averaging means
\begin{eqnarray}
\nonumber
\langle X(\varepsilon_{\bf p})\rangle=\frac{\sum_{{\bf p}}X(\textcolor{black}{\varepsilon_{\bf p}})n_{hh}({\bf p})}{\sum_{{\bf p}}n_{hh}({\bf p})}.
\end{eqnarray}
The massive holes distribution function can be approximated by the Boltzmann distribution, expressed via massive hole density: $n_{hh}({\bf p})=(2\pi P_{hh}/mT)\exp(-\varepsilon_{\bf p}/T)$.

\begin{figure*}[t!]
\includegraphics[width=1.0\columnwidth]{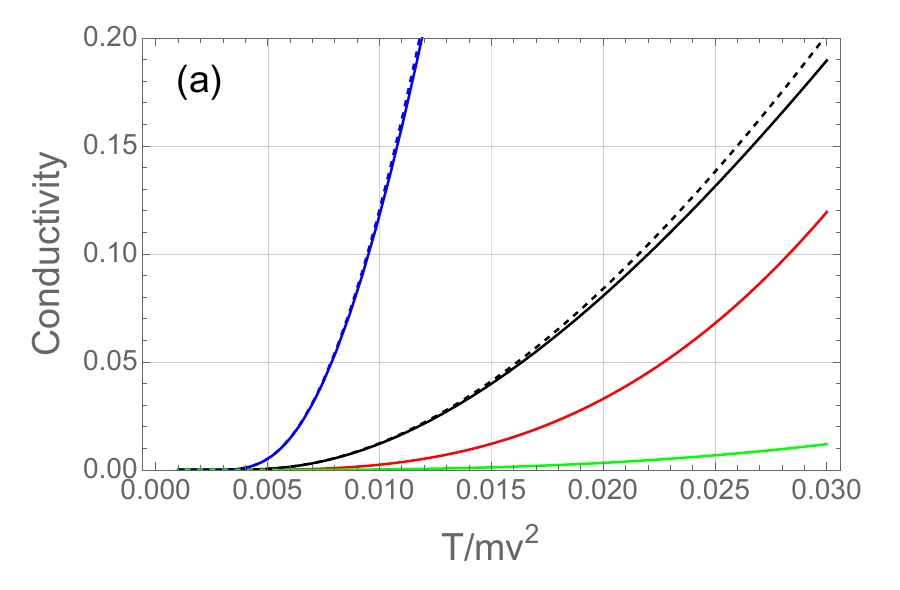} 
\includegraphics[width=0.96\columnwidth]{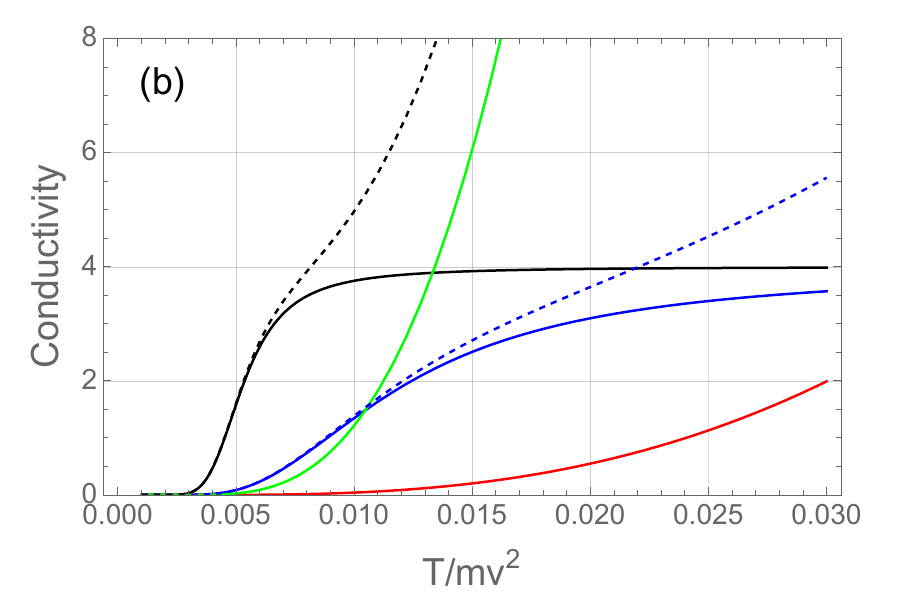} 
\caption{Temperature dependence of Dirac electron (solid blue and black) and massive hole (solid red and green) conductivities for the long-range model in (a) disordered and (b) clean structures. 
Panel (a): blue and red curves correspond to the density of impurities $N_i=5\cdot10^{10}\,cm^{-2}$; whereas black and green curves are for $N_i=5\cdot10^{11}\,cm^{-2}$.
Panel (b): blue and red curves are for $N_i=3\cdot10^{9}\,cm^{-2}$, whereas black and green curves are for $N_i=10^{8}\,cm^{-2}$. 
\textcolor{black}{Dashed curves of the corresponding colors show the total conductivity: $\sigma=\sigma_e+\sigma_{hh}$:
dashed blue curves are the superpositions of solid blue and red curves, while dashed black curves are the superpositions of solid black and green curves.}
The moderate temperature range $T=(50-150)K$ corresponds to $T/mv^2\approx(0.01-0.025)$. 
In calculations, we used the following parameters: $v = 7.0 \times 10^5~\mathrm{m/s}$, 
$m = 0.15\,m_0$, and $\Delta = 15$~meV. 
The conductivities are in units $(e^2/\hbar)(\epsilon\hbar v/2\pi e^2)^2$.}
\label{FigNewLongRange}
\end{figure*}

It should be noted that despite the seeming absence of a direct influence of electron-hole scattering on the massive hole conductivity due to the nearly elastic nature of these processes, the presence of the Dirac electrons component does exert an (indirect) effect on the massive hole conductivity. 
This occurs through the temperature dependence of the massive hole density $P_{hh}$, which is governed by the charge neutrality condition of the system $N_e\approx P_{hh}$, ultimately leading to a nearly quadratic temperature dependence of the hole density~\eqref{HHdensity}.


\section{Temperature behavior of conductivity}

Let us analyze the temperature behavior of the system's conductivity, $\sigma=\sigma_e+\sigma_{hh}$, for different models: short-range and long-range interactions. 

First, the long-range model described in expressions~\eqref{diractimes}, ~\eqref{heavytimes}, and~\eqref{ehtimes} substituted in Eqs.~\eqref{ehcontribution} and~\eqref{hhcontribution} yields
\begin{gather}
\sigma_e=\frac{e^2}{\pi\hbar}\left(\frac{mv^2\tau_C}{\hbar}\right)
\left(\frac{T}{mv^2}\right)^2
\left(\frac{\tau_1/\tau_C}{1+\tau_1/\tau_C}\right)W_0^2(z),\\\nonumber
\sigma_{hh}=\frac{2e^2}{\pi\hbar}\left(\frac{mv^2\tau_C}{\hbar}\right)
\left(\frac{T}{mv^2}\right)^3
W_0^2(z).
\end{gather}
Since the relaxation times satisfy $\tau_C/\tau_1 = P_{hh}/N_i$,  incorporating the temperature dependence of $P_{hh}$ from Eq.~\eqref{HHdensity}, we find $P_{hh}/N_i=\beta(T/mv^2)^2W^2_0(z)$, with $\beta=(mv/\hbar)^2/\pi N_i$ (the factor that controls the influence of impurity scattering on conductivity). 
Then, the conductivities can be presented in the form:
\begin{gather}\label{LongRangeTemp}
\sigma_e=\frac{e^2}{\hbar}\left(\frac{\epsilon\hbar v}{2\pi e^2}\right)^2
\frac{4\beta\left(\frac{T}{mv^2}\right)^2W_0^2(z)}{1+\beta\left(\frac{T}{mv^2}\right)^2W_0^2(z)},\\\nonumber
\sigma_{hh}=\frac{e^2}{\hbar}\left(\frac{\epsilon\hbar v}{2\pi e^2}\right)^2
8\beta\left(\frac{T}{mv^2}\right)^3W_0^2(z).
\end{gather}

Thus, the factor $P_{hh}/N_i$ principally determines the temperature dependence of the conductivity. 
According to Fig.~\ref{Fig3}(c), within the realistic parameter range for HgTe QWs, the massive-hole density in the moderate-temperature regime is of the order of $P_{hh}\sim10^{10}$~cm$^{-2}$. 
Thus, QWs characterized by low impurity densities, \(N_i < P_{hh}\), will be referred to as clean, whereas those with \(N_i > P_{hh}\) are classified as disordered (dirty) QWs. 
In dirty structures, the Dirac electron conductivity scales as $\sigma_e \sim T^2$, whereas it is temperature-independent in clean samples. 

Second, the short-range model described in expressions~\eqref{diractimes}, ~\eqref{heavytimes}, and~\eqref{ehtimes} substituted in Eqs.~\eqref{ehcontribution} and~\eqref{hhcontribution} yields
\begin{gather}
\sigma_e=\frac{e^2}{4\pi\hbar}\left(\frac{mv^2\tau_i}{\hbar}\right)
\left(\frac{\tau_0/\tau_i}{1+\tau_0/\tau_i}\right),\\\nonumber
\sigma_{hh}=\frac{e^2}{\pi\hbar}\left(\frac{mv^2\tau_i}{\hbar}\right)
\left(\frac{T}{mv^2}\right)^2
W_0^2(z).
\end{gather}
In this case, the relaxation times scale as $\tau_{i}/\tau_0 = P_{hh} |U_0|^2/(N_i |u_0|^2)$. 
Substituting the expressions for $\tau_0$ and $\tau_1$ typical for the short-range model, along with the temperature dependence $P_{hh}(T)$, yields the expressions describing the temperature behavior of Dirac electrons and massive holes conductivities:
\begin{gather}\label{ShortRangeTemp}
\sigma_e=\frac{e^2}{\hbar}\left(\frac{\hbar^2}{mu_0}\right)^2
\frac{\beta}{1+\frac{U_0^2}{u_0^2}\beta\left(\frac{T}{mv^2}\right)^2W_0^2(z)},\\\nonumber
\sigma_{hh}=\frac{e^2}{\hbar}\left(\frac{\hbar^2}{mu_0}\right)^2\beta\left(\frac{T}{mv^2}\right)^2W_0^2(z).
\end{gather}
Evidently, in dirty samples, the electronic conductivity is nearly temperature-independent (whereas in clean structures, it scales as $\sigma_e \propto T^{-2}$). 
Moreover, in the moderate temperature regime, the small parameter $T/mv^2 \ll 1$ (where $mv^2 \sim 5 \times 10^3\,\text{K}$ for HgTe QWs) determines the relative weights of the Dirac electron, $\sigma_e$, and massive hole, $\sigma_{hh}$, contributions to the total conductivity, suppressing the massive hole contribution to the total conductivity. 



\section{Results and discussion}
\begin{figure*}[t!]
\includegraphics[width=2.1\columnwidth]{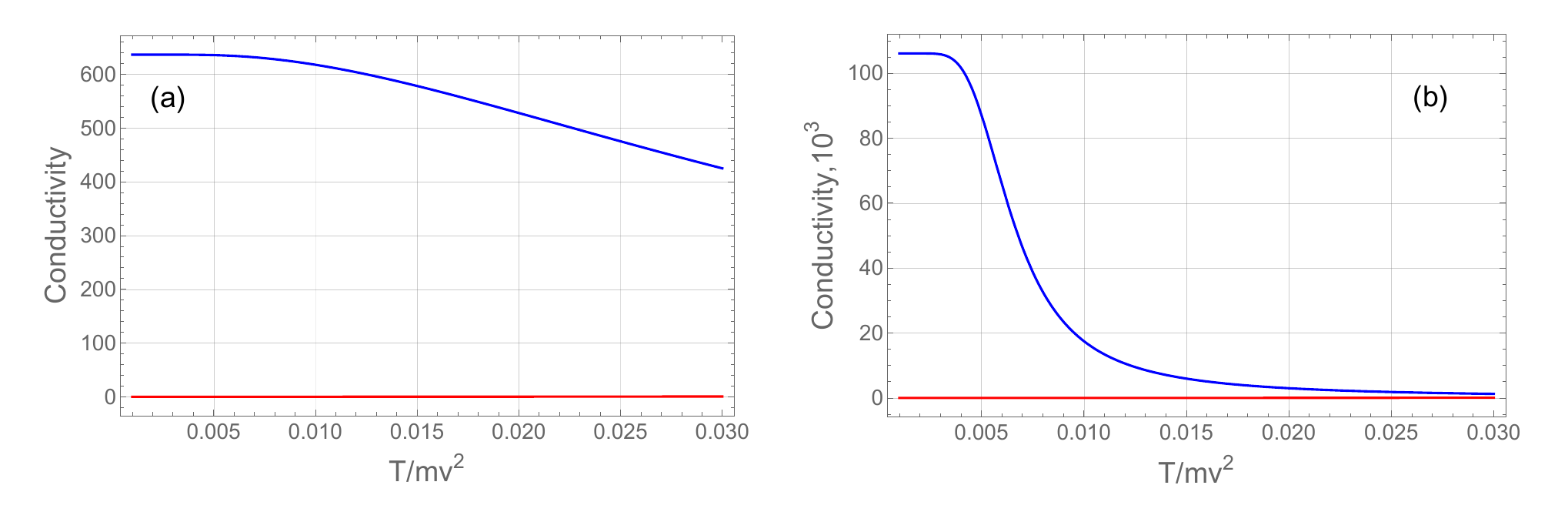} 
\caption{Temperature dependence of Dirac electrons (black) and massive hole (red) conductivities for the short-range model in (a) dirty ($N_i=5\cdot10^{10}$~cm$^{-2}$) and (b) clean ($N_i=3\cdot10^{8}$~cm$^{-2}$) structures. 
The conductivities are in units $(e^2/\hbar)(\hbar^2/mu_0)^2$. 
Due to the smallness of the parameter $(T/mv^2)^2$ in Eq.~\eqref{ShortRangeTemp}, the variation in massive hole conductivity (red) is visually suppressed and appears as a flat line on the scale of Dirac electron conductivity (blue). 
In our calculations, we assume $U_0\approx u_0$ in Eq.~\eqref{ShortRangeTemp} for simplicity.}
\label{FigNewShortRange}
\end{figure*}

\textcolor{black}{
Figure~\ref{FigNewLongRange} and Figure~\ref{FigNewShortRange} show the temperature-dependent conductivities of massless and massive carriers at the charge-neutrality point, alongside the total conductivity ($\sigma=\sigma_e+\sigma_{hh}$). Thus, the analytical expressions given in Eqs.~\eqref{LongRangeTemp} and~\eqref{ShortRangeTemp} are numerically validated. The parameters characterizing the HgTe QW in our model, specifically $v_F$ and $T$, were directly extracted from recent transport experiments on near-critical HgTe QWs~\cite{Levin2026}. 
}

In the moderate-temperature regime, where the Dirac holes' contribution is negligible, the scattering of Dirac electrons by massive holes is nearly elastic, justifying the introduction of a corresponding relaxation time. This yields a complete expression for the conductivity, in which interparticle scattering processes are comparable to particle-impurity scattering mechanisms. The non-trivial density of states of the Dirac component induces characteristic energy dependencies of the relaxation times, ultimately dictating the complex temperature behavior of the total conductivity in this mixture of massive and massless carriers. 

In addition to the relaxation times, the charge-neutrality condition governs the temperature dependence of both the massless electron and massive hole concentrations. Remarkably, the electronic component remains nearly degenerate globally, exhibiting a linear chemical-potential shift with temperature, whereas the massive hole component obeys non-degenerate Boltzmann statistics. This regime allows for an approximate analytical solution to the neutrality equation, yielding the chemical potential and carrier densities as explicit functions of temperature.

\textcolor{black}{The conductivity for the long-range interaction potential demonstrates that the electronic contribution dominates over the holes' contribution in dirty QWs, as shown in Fig.~\ref{FigNewLongRange}(a). In cleaner samples, the holes' contribution competes with its electronic counterpart, as we see in Fig.~\ref{FigNewLongRange}(b), especially with the decrease of the impurity density.}

In the strong screening limit (modeled via contact interactions for both interparticle and impurity scatterings), the massive-hole contribution to conductivity remains negligible across both dirty and clean QWs (Fig.~\ref{FigNewShortRange}(a,b)). In clean samples, the quadratic temperature decay of the electronic conductivity translates into a $T^2$ increase in resistivity. 
This behavior represents a distinctive signature of clean HgTe QWs, contrasting sharply with pristine gapless graphene, where electron-hole scattering of massless particles yields a temperature-independent resistivity~\cite{VyurkovRyzhii, PhysRevB.78.085415, PhysRevB.78.085416}.

Unlike gapped graphene, which typically exhibits an Arrhenius or linear temperature dependence~\cite{PhysRevB.109.085424}, the hybrid system considered here entirely avoids Arrhenius-like behavior even when impurity scattering dominates. Furthermore, while gap opening remains a challenge in graphene~\cite{Schwierz2010}, HgTe QWs offer a highly tunable platform where the QW width determines the bandgap and the gate voltage modulates carrier degeneracy in situ~\cite{doi:10.1126/science.1148047, Snegirev2025, Becker2014}. This structural flexibility enables the study of the transformation from a graphene-like low-temperature state to a moderate-temperature regime dominated by quantum friction between quasirelativistic and non-relativistic particles~\cite{Majumdar2025}. 

\textcolor{black}{Using experimentally established values~\cite{Levin2026} allows us to demonstrate that the transport phenomena predicted by our theory are physically realizable and exhibit good agreement with the behavior observed in these hybrid multi-component plasma systems. Transport in near-critical HgTe systems is dominated by interparticle scattering. Because the linear Dirac spectrum breaks Galilean invariance, the electric current is decoupled from the total momentum, allowing collisions between degenerate Dirac electrons and thermally activated massive holes to directly relax the current without requiring phonons. The strong temperature dependence of the conductivity is driven by the thermal activation of massive holes and the subsequent self-consistent shift of the chemical potential, rather than by standard phonon scattering (which would yield a linear-in-$T$ power law). Such linear behavior was not observed in recent experiments~\cite{Levin2026}, indicating that particle-particle scattering dominates over electron-phonon scattering within this temperature range.}

\textcolor{black}{Let us also summarize the limitations of the model developed in this paper. First, our quantitative predictions are strictly limited to the low-temperature regime, where interparticle and impurity scattering dominate over phonon scattering. 
Another key approximation is the nearly elastic nature of electron-hole scattering processes, which allowed us to disregard hole-electron scattering processes (see Supplemental Material for a detailed derivation~\cite{SMBG}).}


\section*{Conclusions} 
We developed a microscopic theory of interaction-induced conductivity in a two-dimensional mixture of Dirac and massive fermions at the charge-neutrality point. 
Analytical expressions for temperature-dependent conductivity were derived via the linearized Boltzmann transport equation, accounting for intra-species, inter-species, and particle-impurity scattering mechanisms. 
Our results reveal a clear crossover from Dirac-dominated transport at low temperatures to a regime in which interspecies interactions yield a nonconventional, nonmonotonic temperature dependence of the conductivity. 
Quantified for both short-range and long-range interaction potentials, this behavior reflects the onset of quantum friction between distinct fermionic species, a phenomenon uniquely accessible in systems with broken Galilean invariance. \textcolor{black}{A direct comparison between our theory and a recent experiment~\cite{Levin2026} reveals that our model provides a good agreement with the experimental data.
}

Our analysis highlights the critical role of the charge neutrality point as a natural setting in which the chemical potential is self-consistently determined by temperature, eliminating external-doping complexities and enabling a clean, intrinsic probe of interaction-driven effects. 
The predicted temperature dependence of the conductivity provides clear experimental signatures that were recently tested in high-mobility HgTe quantum wells.

This work establishes HgTe-based heterostructures at charge neutrality as a versatile and controllable platform for exploring non-equilibrium many-body physics. 
The ability to independently tune the band gap, carrier degeneracy, and scattering channels offers a unique advantage over graphene for isolating fundamental interaction-driven transport phenomena. 
Our findings thus lay the groundwork for future experimental investigations into quantum friction, hydrodynamic transport, and emergent collective behavior in two-dimensional systems where multiple carrier species coexist and interact. 
This platform holds promise not only for advancing fundamental understanding but also for potential applications in low-power, interaction-controlled electronic devices operating in the partially non-degenerate regime.

\section*{Acknowledgements} 
We were supported by the National Natural Science Foundation of China (NSFC) under Grant No.~W2532001, Guangdong Basic and Applied Basic Research Foundation under Grant No.~2026A1515012415,
the Ministry of Science and Higher Education of the Russian Federation (Project FSUN-2026-0004), and the Foundation for the Advancement of Theoretical Physics and Mathematics ``BASIS''. 

\bibliography{biblio}
\bibliographystyle{apsrev4-2}

\end{document}


\title{Supplemental Material for \textit{Conductivity of charge-neutral multicomponent 2D electron-hole system}}

\author{Yuping Huang}
\affiliation{Department of Materials Science and Engineering, Southern University of Science and Technology, Shenzhen, 518055, China}
\affiliation{Guangdong Technion -- Israel Institute of Technology, Shantou, Guangdong, 515063, China}

\author{O.~V.~Kibis}
\affiliation{Novosibirsk State Technical University, Novosibirsk 630073, Russia}

\author{V.~M.~Kovalev}
\affiliation{Novosibirsk State Technical University, Novosibirsk 630073, Russia}

\author{I.~G.~Savenko}
\affiliation{Guangdong Technion -- Israel Institute of Technology, Shantou, Guangdong, 515063, China}

\date{\today}

\maketitle

\tableofcontents

\section{Screening effects}
\label{LabSec01}

Here, we investigate the screening effect and its impact on the scattering times of Dirac electrons. The primary contribution to screening originates from the gate electrode inherent in realistic transport measurement configurations, whereas the secondary contribution arises from the collective screening by mobile carriers. In the former case, the screening yields a Coulomb potential, whose Fourier transform is given by
\begin{equation}
U_{\bf q}=\frac{2\pi e^2}{\epsilon q}\left(1-e^{-qd}\right),
\end{equation}
where $\mathbf{q}$ is the transferred momentum, and $d$ denotes the distance between the quantum well (QW) hosting the mobile carriers and the metallic gate. 

The mobile carriers introduce an additional screening factor, which can be taken into account within the standard random phase approximation in the static limit as:
\begin{equation}
U_{\bf q}\rightarrow U_{\bf q}^s=\frac{U_{\bf q}}{1-U_{\bf q}\Pi_{\bf q}}=
\frac{2\pi e^2}{\epsilon}\frac{\left(1-e^{-qd}\right)}{q+q_s\left(1-e^{-qd}\right)};\,\,\,\,q_s=\frac{2\pi e^2}{\epsilon}\frac{\partial N}{\partial \mu},
\end{equation}
where \(q_{s}\) is the screening wave vector, and \(N = N_e + P_{hh} = 2N_e=2\mu^2/(4\pi v^2)\) is the total particle density in the moderate-temperature range with heavy holes dominating over the Dirac hole density. 
A specific feature of this expression is that the chemical potential, and consequently the screening wave vector, depend on temperature. 
Further analysis can be performed within the framework of our analytical model.

The momentum relaxation times of Dirac particles are given by
\begin{equation}
\begin{pmatrix}
\dfrac{1}{\tau_e(\varepsilon_{\bf k})} \\[0.5ex]
\dfrac{1}{\tau_{eh}(\varepsilon_{\bf k})} 
\end{pmatrix}=
2\pi\begin{pmatrix}
N_i \\
P_{hh}
\end{pmatrix}
\sum_{\mathbf{k'}}
|U^s_{\mathbf{k}'-\mathbf{k}}|^2{\cal F}_{\bf kk'}(1-\cos\theta_{\mathbf{k}'\mathbf{k}})\delta(\varepsilon_{{\bf k}'}-\varepsilon_{\bf k}).
\end{equation}
Here, $\mathbf{k}$ is the initial momentum of the Dirac particle, and the factor ${\cal F}_{\mathbf{k}, \mathbf{k}+\mathbf{q}} = \cos^2(\theta/2)$ accounts for the angular dependence of the wavefunctions of Dirac particles, where $\theta$ is the scattering angle.
A general integration here is cumbersome and cannot be performed analytically. 
Thus, it is instructive to consider two limiting cases that elucidate the influence of the screening effects and the resulting temperature dependences of the respective momentum relaxation times. 
The first case corresponds to a distant gate (\(d \rightarrow \infty\)), whereas the second corresponds to a gate positioned close to the conducting QW (\(d \rightarrow 0\)).


\subsection{Distant gate}
In the former case $d\rightarrow\infty$, we find
\begin{equation}
\begin{pmatrix}
\dfrac{1}{\tau_e(\varepsilon_{\bf k})} \\[0.5ex]
\dfrac{1}{\tau_{eh}(\varepsilon_{\bf k})} 
\end{pmatrix}=
2\pi\begin{pmatrix}
N_i \\
P_{hh}
\end{pmatrix}
\sum_{\mathbf{k'}}
\left|\frac{2\pi e^2}{\epsilon(q_s+|\mathbf{k}'-\mathbf{k}|)}\right|^2
{\cal F}_{\bf kk'}
(1-\cos\theta_{\mathbf{k}'\mathbf{k}})\delta(\varepsilon_{{\bf k}'}-\varepsilon_{\bf k}).
\end{equation}
This expression can be integrated analytically, yielding
\begin{equation}
\begin{pmatrix}
\dfrac{1}{\tau_e(\varepsilon_{\bf k})} \\[0.5ex]
\dfrac{1}{\tau_{eh}(\varepsilon_{\bf k})} 
\end{pmatrix}=
\frac{1}{\pi\varepsilon_{\bf k}}\begin{pmatrix}
N_i \\
P_{hh}
\end{pmatrix}
\left(\frac{2\pi e^2}{\epsilon}\right)^2\int\limits_0^1dt\frac{t^2\sqrt{1-t^2}}{(t+\eta)^2},\,\,\,\,\eta=\alpha_g\frac{\mu(T)}{2\varepsilon_{\bf k}},
\end{equation}
with $\alpha_g=e^2/(\epsilon\hbar v)$ the effective fine-structure constant introduced for Dirac particles in graphene. 
For HgTe quantum wells $v\approx 7\times10^7$~cm/s and $\epsilon=20\div50$ we estimate $\alpha_g=0.16\div0.06$. 
Consequently, the parameter \(\eta \), which depends on the Dirac particle energy, introduces an additional temperature dependence via \(\mu(T)\). 
When substituted into the expression for conductivity, these momentum relaxation times are evaluated at \(\varepsilon_{\mathbf{k}} = \mu\) [see Eq. (21) in the manuscript]. Thus, \(\eta \approx \alpha_g \ll 1\), indicating that screening plays a negligible role, and these relaxation time expressions reduce directly to the long-range model considered in the manuscript.


\subsection{Gate close to QW}
In the second case $d\rightarrow0$, we find
\begin{equation}
\begin{pmatrix}
\dfrac{1}{\tau_e(\varepsilon_{\bf k})} \\[0.5ex]
\dfrac{1}{\tau_{eh}(\varepsilon_{\bf k})} 
\end{pmatrix}=
2\pi\begin{pmatrix}
N_i \\
P_{hh}
\end{pmatrix}
\left(\frac{2\pi e^2d}{\epsilon(1+q_sd)}\right)^2
\sum_{\mathbf{k'}}
{\cal F}_{\bf kk'}
(1-\cos\theta_{\mathbf{k}'\mathbf{k}})\delta(\varepsilon_{{\bf k}'}-\varepsilon_{\bf k}).
\end{equation}
This expression directly corresponds to our short-range model with identical parameters $|u_0|=|U_0|=2\pi e^2/\epsilon(1+q_sd)$, where (restore $\hbar$ for dimensionality) 
\begin{equation}
q_sd=\frac{2\pi e^2d}{\epsilon}\frac{\mu(T)}{\pi v^2}=\alpha_g\frac{2\mu(T)d}{\hbar v}=
4\alpha_g\left(\frac{mvd}{\hbar}\right)\left(\frac{T}{mv^2}\right)W_0(z). 
\end{equation}
Due to the smallness of \(\alpha_g \sim 0.1\) and \(T / m v^2 \sim 10^{-2}\), $W(z)\sim1$ (at the largest temperature $T\sim100$~K), the parameter \(q_s d \ll1\), and any potential temperature dependence of the parameters \(u_{0}\) and \(U_{0}\) plays a negligible role in the analysis of Dirac electrons' conductivity in a moderate temperature regime.

Consequently, direct evaluations of screening effects indicate that they either have a negligible impact on relaxation times (when the gate is distant) or reduce to a short-range interaction model (when the gate is positioned close to the QW). 
Thus, our theoretical analysis, based on models of short- and long-range interaction potentials, directly corresponds to actual experimental configurations.


\section{Estimation of the $Q_{he}$ correction}
\label{LabSec02}

The linearized collision integral, describing the change of the nonequilibrium distribution function of holes due to their collisions with electrons, reads as:
\begin{equation}
\label{EqQhe01}
Q_{he}(\mathbf{p}) = \frac{2\pi}{\hbar} 
\int d\omega
\int \frac{d^2k}{(2\pi\hbar)^2} 
\int \frac{d^2q}{(2\pi\hbar)^2} |U_q|^2 
{\cal F}_{\mathbf{k}, \mathbf{k}+\mathbf{q}}  
\mathcal{D}(\mathbf{p}, \mathbf{k}, \mathbf{q}) \delta(\epsilon_{\mathbf{k}+\mathbf{q}} - \epsilon_{\mathbf{k}} - \omega) 
\delta(\epsilon_{\mathbf{p}-\mathbf{q}} - \epsilon_{\mathbf{p}} + \omega),
\end{equation}
where $\mathbf{p}$ is the initial momentum of the hole and $\omega$ is the transferred energy. 
To accurately describe the relaxation kinetics, the thermodynamic driving factor must strictly mirror `in - out' scattering transitions:
\begin{equation}
\label{EqD01}
\mathcal{D}(\mathbf{p}, \mathbf{k}, \mathbf{q}) = 
\frac{[n_{hh}(\mathbf{p}-\mathbf{q}) - n_{hh}(\mathbf{p})][n_e(\mathbf{k}+\mathbf{q}) - n_e(\mathbf{k})]}{(-4T)\sinh^2\left(\frac{\omega}{2T}\right)} \Big[ \chi_h(\mathbf{p}-\mathbf{q}) - \chi_h(\mathbf{p}) + \chi_e(\mathbf{k}+\mathbf{q}) - \chi_e(\mathbf{k}) \Big].
\end{equation}

Due to the energy-conserving delta functions in~\eqref{EqQhe01}, we express the distribution differences via their energy expansions. For small energy transfers $\omega \ll T, \mu$, which is the case in the manuscript, we expand:
\begin{equation}
\label{EqExpansionDens01}
n_{hh}(\mathbf{p}-\mathbf{q}) - n_{hh}(\mathbf{p})
\approx -\frac{\partial n_{hh}(\varepsilon_p)}{\partial \varepsilon_p}\omega, \quad
n_{e}(\mathbf{k}+\mathbf{q}) - n_{e}(\mathbf{k})
\approx 
\frac{\partial n_{e}(\varepsilon_k)}{\partial \varepsilon_k}\omega.
\end{equation}
Using the property of the Maxwellian distribution, $\frac{\partial n_{hh}}{\partial \varepsilon_p} = -n_{hh}(\mathbf{p})/T$, the product of the differences in the numerator becomes $-n_{hh}(\mathbf{p}) \frac{\partial n_e}{\partial \varepsilon_k} \omega^2/T$. Expanding the denominator as $-4T\sinh^2\left(\frac{\omega}{2T}\right) \approx - \omega^2/T$, the explicit dependence on the energy transfer algebraically cancels out completely:
\begin{equation}
\frac{- \frac{1}{T} n_{hh}(\mathbf{p}) \frac{\partial n_e}{\partial \varepsilon_k} \omega^2}{- \omega^2 / T} = n_{hh}(\mathbf{p}) \frac{\partial n_e}{\partial \varepsilon_k}.
\end{equation}
This rigorous cancellation avoids any artificial singularities. 

Furthermore, the term $\mathcal{D}$ simplifies to:
\begin{equation}
\mathcal{D} \approx n_{hh}(\mathbf{p}) \frac{\partial n_e}{\partial \varepsilon_k} \Big[ \chi_h(\mathbf{p}-\mathbf{q}) - \chi_h(\mathbf{p}) + \Delta \chi_e \Big],
\end{equation}
where $\Delta \chi_e = \chi_e(\mathbf{k}+\mathbf{q}) - \chi_e(\mathbf{k})$. 

Since the hole mass $m_h \gg m_e$, the transferred momentum $q \ll p$. 
We can expand the hole perturbation to first order, naturally generating a minus sign characteristic of a restorative force:
\begin{equation}
\chi_h(\mathbf{p}-\mathbf{q}) - \chi_h(\mathbf{p}) \approx - q_j \frac{\partial \chi_h}{\partial p_j}.
\end{equation}
Substituting $\mathcal{D}$ into~\eqref{EqQhe01}, the integration over $d\omega$ consumes the first delta function, setting $\omega = \epsilon_{\mathbf{k}+\mathbf{q}} - \epsilon_{\mathbf{k}} \equiv \Delta\epsilon_e$. The remaining delta function enforces global energy conservation: $\delta(\Delta\epsilon_e - \mathbf{v}_h \cdot \mathbf{q} + \frac{q^2}{2m})$. 
Expanding to the first order for small velocities gives:
\begin{equation}
\delta\left(\Delta\epsilon_e - \frac{\mathbf{p}\cdot\mathbf{q}}{m}\right) \approx \delta(\Delta\epsilon_e) - \frac{p_i q_i}{m} \delta'(\Delta\epsilon_e).
\end{equation}
Isolating the term representing the diffusion of holes due to the equilibrium electron background ($\chi_e = 0$) yields:
\begin{equation}
Q_{he}^{\textrm{(sc)}}(\mathbf{p}) = \frac{2\pi}{\hbar} n_{hh}(\mathbf{p}) \frac{\partial \chi_h}{\partial p_j} \int \frac{d^2k}{(2\pi\hbar)^2} \int \frac{d^2q}{(2\pi\hbar)^2} |U_q|^2 {\cal F}_{\bf kk'}
\left(-q_j\right) \frac{\partial n_e}{\partial \varepsilon_k} \left[ \delta(\Delta\epsilon_e) - \frac{p_i q_i}{m} \delta'(\Delta\epsilon_e) \right].
\end{equation}
The first term in the brackets is odd with respect to the scattering axis and vanishes. 
The cross-product of the two minus signs (from the perturbation gradient and the delta expansion) naturally generates the surviving rank-2 diffusion tensor with a purely restorative (negative) overall geometry:
\begin{equation}\label{scattering}
Q_{he}^{\textrm{(sc)}}(\mathbf{p}) = - n_{hh}(\mathbf{p}) \frac{p_i}{m} \frac{\partial \chi_h}{\partial p_j} \cdot \frac{2\pi}{\hbar} \int \frac{d^2k}{(2\pi\hbar)^2} \int \frac{d^2q}{(2\pi\hbar)^2} |U_0|^2 {\cal F}_{\bf kk'}
q_i q_j \frac{\partial n_e}{\partial \varepsilon_k} \delta'(\epsilon_{\mathbf{k}'} - \epsilon_{\mathbf{k}}).
\end{equation}
%


\subsection{Evaluating the integral}

Switching to polar coordinates and aligning the initial electron momentum $\mathbf{k}$, we map $d^2k = \frac{\varepsilon}{v^2} d\varepsilon d\phi_k$ and $d^2{k'} = \frac{\varepsilon'}{v^2} d\varepsilon' d\theta$. The angular integration of $q_i q_j$ over the initial orientation $\phi_k$ yields a Kronecker delta:
\begin{equation}
\int_0^{2\pi} q_i q_j d\phi_k = \pi q^2 \delta_{ij} = \frac{\pi}{v^2} \left( \varepsilon^2 + \varepsilon'^2 - 2\varepsilon\varepsilon'\cos\theta \right) \delta_{ij}.
\end{equation}
We integrate over the final electron energy $\varepsilon'$ utilizing the derivative of the delta function, $\int f(\varepsilon') \delta'(\varepsilon' - \varepsilon) d\varepsilon' = -f'(\varepsilon)$:
\begin{equation}
\int d\varepsilon' \frac{\varepsilon'}{v^2} \frac{\pi}{v^2} \left( \varepsilon^2 + \varepsilon'^2 - 2\varepsilon\varepsilon'\cos\theta \right) \delta'(\varepsilon' - \varepsilon) = - \frac{4\pi \varepsilon^2}{v^4}(1 - \cos\theta) = - \frac{8\pi \varepsilon^2}{v^4} \sin^2\left(\frac{\theta}{2}\right).
\end{equation}
Next, performing the radial integration over the initial electron energy $\varepsilon$, the derivative of the Fermi function for degenerate Dirac electrons acts as $- \frac{\partial n_e}{\partial \varepsilon} \approx \delta(\varepsilon - \mu)$. The double-negative combination makes this physical space integration positive:
\begin{equation}
- \frac{8\pi}{v^4} \sin^2\left(\frac{\theta}{2}\right) \int_0^\infty \frac{\varepsilon}{v^2} \frac{\partial n_e}{\partial \varepsilon} \varepsilon^2 d\varepsilon = \frac{8\pi \mu^3}{v^6} \sin^2\left(\frac{\theta}{2}\right).
\end{equation}
Finally, we integrate over the scattering angle $\theta$ with the Dirac overlap factor $F_{\mathbf{k}, \mathbf{k}'} = \cos^2(\theta/2)$:
\begin{equation}
\int_0^{2\pi} \cos^2\left(\frac{\theta}{2}\right) \sin^2\left(\frac{\theta}{2}\right) d\theta = \frac{\pi}{4}.
\end{equation}
Assembling all numerical factors and returning to the tensor expression gives:
\begin{equation}
Q_{he}^{\textrm{(sc)}}(\mathbf{p}) = - n_{hh}(\mathbf{p}) \frac{p_i}{m} \frac{\partial \chi_h}{\partial p_j} \delta_{ij} \left[ \frac{2\pi}{\hbar} \frac{1}{(2\pi\hbar)^4} |U_0|^2 \frac{8\pi \mu^3}{v^6} \frac{\pi}{4} \right] = - n_{hh}(\mathbf{p}) \left( \textbf{p}\cdot\frac{\partial \chi_h}{\partial \textbf{p}} \right) \frac{|U_0|^2 \mu^3}{4\pi m v^6 \hbar^5}.
\end{equation}

The actual distribution perturbation relates to the energy shift function as $\delta f_h = -\frac{\partial n_{hh}}{\partial \varepsilon_p} \chi_h = \frac{n_{hh}}{T} \chi_h$. 
Assuming linear response $\chi_h \propto p$, we deduce $\textbf{p}\cdot \frac{\partial \chi_h}{\partial \textbf{p}} = \chi_h = \frac{T}{n_{hh}} \delta f_h$. 
The scalar relaxation frequency, defined via $Q_{he}^{\textrm{(sc)}} = - \nu_{he} \delta f_h$, then reads as:
\begin{equation}
\nu_{he} = \frac{T |U_0|^2 \mu^3}{4\pi \hbar^5 m v^6}.
\end{equation}
%


\subsection{Dynamic friction: electron drag contribution}

We now evaluate the contribution to the collision integral arising from the nonequilibrium electron perturbation, $\Delta \chi_e$, which acts as dynamic friction (drag) exerted on the holes:
\begin{equation}
Q_{he}^\textrm{(dr)}(\mathbf{p}) = \frac{2\pi}{\hbar} n_{hh}(\mathbf{p}) \int \frac{d^2k}{(2\pi\hbar)^2} \int \frac{d^2q}{(2\pi\hbar)^2} |U_0|^2 {\cal F}_{\bf kk'}
\frac{\partial n_e}{\partial \varepsilon_k} \Delta\chi_e \left[ \delta(\Delta\epsilon_e) - \frac{p_i q_i}{m} \delta'(\Delta\epsilon_e) \right].
\end{equation}
The nonequilibrium electron perturbation is given by $\chi_e(\mathbf{k}) = \tau_e v (\mathbf{F}\cdot\mathbf{n}_k) = \frac{\tau_e v^2}{\mu} \mathbf{F} \cdot \mathbf{k}$. Since scattering is peaked on the Fermi surface ($k \approx p_F = \mu/v$), the transferred difference is:
\begin{equation}
\Delta \chi_e = \chi_e(\mathbf{k}+\mathbf{q}) - \chi_e(\mathbf{k}) \approx \frac{\tau_e v^2}{\mu} F_j q_j.
\end{equation}
Substituting this into the integral, the first term inside the brackets contains an odd power of $q$ and vanishes angularly. Coupling with the $\delta'$ expansion brings one explicit negative sign, canceling the negative sign generated inside the $\delta'$ integration space:
\begin{equation}\label{drag}
Q_{he}^\textrm{(dr)}(\mathbf{p}) = - n_{hh}(\mathbf{p}) \frac{p_i}{m} \left( \frac{\tau_e v^2}{\mu} F_j \right)  \frac{2\pi}{\hbar} \int \frac{d^2k}{(2\pi\hbar)^2} \int \frac{d^2q}{(2\pi\hbar)^2} |U_0|^2 {\cal F}_{\bf kk'}
\frac{\partial n_e}{\partial \varepsilon_k} q_i q_j \delta'(\epsilon_{\mathbf{k}'} - \epsilon_{\mathbf{k}}).
\end{equation}
Substituting the evaluated integral  from the previous section gives:
\begin{equation}
Q_{he}^\textrm{(dr)}(\mathbf{p}) = n_{hh}(\mathbf{p}) 
\frac{\textbf{p}\cdot\textbf{F}}{m} \left( \frac{\tau_e v^2}{\mu} \right) \left[ \frac{|U_0|^2 \mu^3}{4\pi \hbar^5 v^6} \right] = n_{hh}(\mathbf{p}) (\textbf{p}\cdot\textbf{F}) \frac{|U_0|^2 \tau_e \mu^2}{4\pi m v^4 \hbar^5}.
\end{equation}
This yields a driving drag force.


\subsection{Comparison with impurity scattering of holes}

To confirm the physical validity of neglecting the electron-hole collision integral $Q_{he} = Q_{he}^{\textrm{(sc)}} + Q_{he}^\textrm{(dr)}$ in the transport equations, we should compare the electron-hole scattering frequency $\nu_{he}$ with the hole-impurity scattering frequency $\nu_{hi}$, and the dynamic friction (drag) term $Q_{he}^\textrm{(dr)}$ with the primary external driving force.

First, the scattering frequency of heavy holes on a short-range impurity potential $u_0$ with concentration $N_i$ equals:
\begin{equation}
\nu_{hi} = \frac{m N_i |u_0|^2}{\hbar^3}.
\end{equation}
Using the 2D Dirac gas relation for the electron density $N_e = \mu^2 / (4\pi \hbar^2 v^2)$, we find
\begin{gather}
\frac{Q_{he}^\textrm{(sc)}}{Q_{hi}}=\frac{\nu_{he}}{\nu_{hi}} \sim \left( \frac{U_0}{u_0} \right)^2 \frac{T \mu^3}{4\pi \hbar^2 m^2 v^6 N_i} \sim \left( \frac{N_e}{N_i} \right)  \left( \frac{U_0}{u_0} \right)^2  \left( \frac{\mu}{m v^2} \right)^2  \left( \frac{T}{\mu} \right)\\
\nonumber
=
\left( \frac{N_e}{N_i} \right)  \left( \frac{U_0}{u_0} \right)^2  \left( \frac{T}{m v^2} \right)^2  \left( \frac{\mu}{T} \right)\ll1.
\end{gather}
This inequality originates from the relations $N_e=P_{hh}\sim (0.1\div10)N_i$, $U_0\approx u_0$ (see Sec.~\ref{LabSec01}), $T/mv^2\sim10^{-2}$ and $\mu/T\sim1\div10$.

Furthermore, to estimate the drag contribution, $Q_{he}^\textrm{(dr)}$, we use Eqs.~\eqref{scattering} and~\eqref{drag}, yielding:
\begin{equation}
\frac{Q_{he}^\textrm{(dr)}}{Q_{he}^\textrm{(sc)}}=
\nu\tau_e\frac{mv^2}{\mu}
=\frac{N_e}{P_{hh}}\frac{T}{mv^2}\ll1,
\end{equation}
where, we employ the relations $\tau_e^{-1} = P_{hh} U_0^2 m / \hbar^3$ 
and $N_e = P_{hh}$, which follow directly from the charge neutrality condition. 
Consequently, the hierarchical relation $Q_{he}^\textrm{(dr)} \ll Q_{he}^\textrm{(sc)} \ll Q_{hi}$ holds. 
This establishes that, owing to the extreme kinematics of momentum mismatch, 
hole scattering by impurities dominates over both the 
relaxation and drag contributions of electron-hole scattering. 
This separation of scales rigorously justifies the approximation $Q_{he} \approx 0$ 
across standard operational regimes we considered.